\documentclass[12pt,a4paper,notitlepage]{article}

\usepackage{amsmath,amssymb}

\usepackage[dvips]{graphicx}

\begin{document}

\title{$\Theta^+$ and $\Lambda(1520)$ production in pp reactions at high energy}

%\classification{13.60.Rj 13.75.Jz 13.85.Ni} \keywords
%{Pentaquarks, excited hyperons, inclusive production}

\author{I.M.Narodetskii and M.A.Trusov\\[4mm] \itshape{ITEP, Moscow, Russia}}

\maketitle

\begin{abstract}
We estimate the  cross sections for the inclusive production of
$\Theta^+$ and $\Lambda(1520)$ in $pp$ collisions at high energy
using the $K$ exchange diagrams. We find that inclusive $\Theta^+$
production should be at the level of 1 $\mu$b at energies~
$\sqrt{s}~\gtrsim~10~{\rm GeV}$. The ratio of $\Theta^+(1540)$ to
$\Lambda(1520)$ production cross-sections is $\sim 1\%$.
\end{abstract}

%\newpage

%%%%%%%%%%%%%%%%%%%%%%%%%%%%%%%%%%%%%%%%%%%%
%% MAINMATTER
%%%%%%%%%%%%%%%%%%%%%%%%%%%%%%%%%%%%%%%%%%%%
\section{Introduction}
The possible existence of the $\Theta^+$ pentaquark remains one of
the puzzling mysteries  of recent years. To date there are more
than 20 experiments with evidence for this state, but criticism
for the $\Theta^+$ claim arises because similar number of high
energy experiments did not find any evidence for the $\Theta^+$,
even though the ``conventional'' three-quark hyperons states, such
as $\Lambda(1520)$ resonance, are seen clearly\footnote{ The
situation is getting more intriguing, as recently CLAS
collaboration reported negative results on $\Theta^+$
photoproduction off proton and deuteron with high statistics.
Meanwhile LEPS collaboration reported the new evidence of the
$\Theta^+$ in $\gamma d\to\Theta^+\Lambda^*(1520)$. Also, STAR
collaboration observed the doubly charged exotic baryon in the
$pK^+$ decay channel. For
 a recent review see \cite{jlab}.} . The superposition of positive and
 negative results is very disturbing. New experiments are needed
 to confirm or refute pentaquark existence.

Most of negative high energy experiments are high statistic hadron
beam experiments. {\it E.g.} HERA-B, a fixed target experiment at
the 920 GeV proton storage ring of DESY \cite{HERA-B} finds no
evidence for narrow signals in the $\bar K^0p$ channel and only
sets the upper limit on production $\Theta^+$ and $\Lambda(1520)$
cross sections in mid-rapidity region. This negative result would
present serious rebuttal evidence to worry about. However, without
obvious production mechanism of the $\Theta^+$ (if it exists) or
even $\Lambda(1520)$ the rebuttal is not very convincing. As of
today, the only positive high-energy hadron beam result is that by
SVD Collaboration \cite{SVD} who finds a narrow $K_S^0p$ resonance
in the inclusive $pA\to \Theta^+X$ reaction using  70 GeV proton
beam at IHEP accelerator.

In this paper we estimate the high-energy behavior of $\Theta^+$
and $\Lambda(1520)$ production  in inclusive $pp$ processes in the
fragmentation region using the $K$-exchange diagrams. Our
conclusion is that the cross section of the $\Theta^+$-production
is suppressed compared to the production of  $\Lambda(1520)$ by a
factor of $\sim10^{-2}$. This suppression is mainly due to
smallness of the coupling constant $G^2_{\Theta KN}$ compared  to
$G^2_{\Lambda KN}$ that in turn is related to the small width of
the $\Theta^+$.

\section{Inclusive production cross sections}

\begin{figure}
\begin{center}
\includegraphics[width=40mm,keepaspectratio=true]{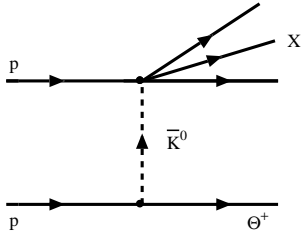}
\end{center}
\caption{$\Theta^+$ production in inclusive pp scattering}
\label{fig:pp_inclusive}
\end{figure}

The ${\bar K^0}$-exchange diagram for $pp\to\Theta^+X$ is shown in
fig. \ref{fig:pp_inclusive}. In the high energy limit the
corresponding cross section written in terms of the Feynman
variable $x_F$, the fraction of incident proton momentum carried
by the $\Theta^+$ (in the center-of-mass system), and $k_{\bot}$,
the transverse momentum of the $\Theta^+$ relative to the initial
proton momentum,  reads
\begin{equation}
\label{dsigma} \frac{d\sigma}{dx_Fdk_{\bot}^2}
=\frac{1}{4\pi}\cdot\frac{G_{\Theta KN}^2}{4\pi}\cdot
\frac{1-x_F}{x_F}\cdot \Phi_{\Theta}(t)\cdot
F^4(t)\cdot\sigma_{{\bar K}^0N}(s_1),
\end{equation}
where $s_1=(1-x_F)s$, $s=4(p^2+m_p^2)$ is total center-of-mass
energy squared, $F(t)$ is the phenomenological form factor, $t$ is
the 4-momentum transfer squared, and $\Phi_{\Theta}(t)$ is the
squared product of the vertex function for $p\to\Theta^+\bar K^0$
and kaon propagator:
\begin{equation}
\label{phi}
\Phi_{\Theta}(t)=\frac{(m_N-m_{\Theta})^2-t}{(t-m_K^2)^2}.
\end{equation}
At high energy \begin{equation}t\approx
m_{\Theta}^2+m_p^2(1-x_F)-\frac{m_{\Theta}^2+k_{\bot}^2}{x_F},
\end{equation} and the total cross section is energy independent.

The cross section for the inclusive $\Lambda(1520)$ production has
the similar form with the substitutions
%\begin{center}
%\end{center}
\begin{equation}
G_{\Theta KN}\to G_{\Lambda KN},~~~\Phi_{\Theta}\to\Phi_{\Lambda}
=\frac{(m_p+m_{\Lambda})^2-t}{6m_{\Lambda}^2m_K^2}\cdot
\frac{((m_p-m_{\Lambda})^2-t)^2}{(t-m_K^2)^2},
\end{equation}
\subsection{The $KN\Theta^+$ vertex}
 The
$KN\Theta^+$ vertex for $J^P(\Theta^+)=\frac{1}{2}^+$ is
\begin{equation}
\label{NKT} { L}_{\Theta KN}=iG_{\Theta
KN}(K^{\dag}{\bar\Theta}\gamma_5N+\bar N\gamma_5\Theta K).
%\nonumber
\end{equation} The Lagrangian of Eq.
(\ref{NKT}) corresponds with the $\Theta^+$ being a $p$-wave
resonance in the $K^0p$ system. The partial decay width
$\Gamma_{\Theta\to K^0p}$ is
\begin{equation}\Gamma_{\Theta\to
K^0p}=\frac{G^2_{\Theta
KN}}{4\pi}\cdot\frac{2p_K^3}{(m_{\Theta}+m_p)^2-m_K^2} ,
%\nonumber
\end{equation} where $p_K=260$ MeV/c is the kaon
momentum in the rest frame of $\Theta^+$. The coupling constant
$G_{KN\Theta}$  can be expressed through the $\Theta^+$ width as
%Using the reference value of width
%$\Gamma_{\Theta\to K^0p}}=1~{\rm MeV}$ one has
\begin{equation}
\frac{G_{\Theta KN}^2}{4\pi}=0.167\cdot\frac{ \Gamma_{\Theta\to
K^0p}}{1~{\rm MeV}}.
%\nonumber
\end{equation}
\subsection{$KN\Lambda(1520)$ vertex}
The $KN\Lambda(1520)$ vertex is
\begin{equation}
\label{KNL} L_{\Lambda KN}=\frac{G_{\Lambda KN}}{m_K}\left(\bar
\Lambda^{\mu}\gamma_5N\partial_{\mu}K+\bar
N\gamma_5\Lambda^{\mu}\partial_{\mu}K^{\dag}\right),
%\nonumber
\end{equation}
where $\Lambda^{\mu}$ is the vector spinor  for the spin 3/2
particle. The Lagrangian of Eq. (\ref{KNL}) corresponds with the
$\Lambda(1520)$ being a $d$-wave resonance in the $K^-p$ system.
The  $\Lambda(1520)\to pK^-$ width is
\begin{equation}\Gamma_{\Lambda^*\to K^-p}=\frac{G_{\Lambda KN}^2}{4\pi}\cdot
\frac{2p_K^5}{3m_K^2}\cdot\frac{1}{(m_{\Lambda}+m_p)^2-m_K^2},
%\nonumber
\end{equation} where $p_K=246~{\rm MeV/c}$ is the kaon
momentum in the rest frame of $\Lambda(1520)$. Using the PDG
values of $\Gamma_{tot}(\Lambda(1520))=15.6$ MeV and
$BR(\Lambda(1520)\to NK)=45\%$ we obtain
\begin{equation}
\frac{G_{\Lambda KN}^2}{4\pi}\approx 8.14.
\end{equation}
\section{Results}
\begin{figure}
\includegraphics[width=60mm,keepaspectratio=true]{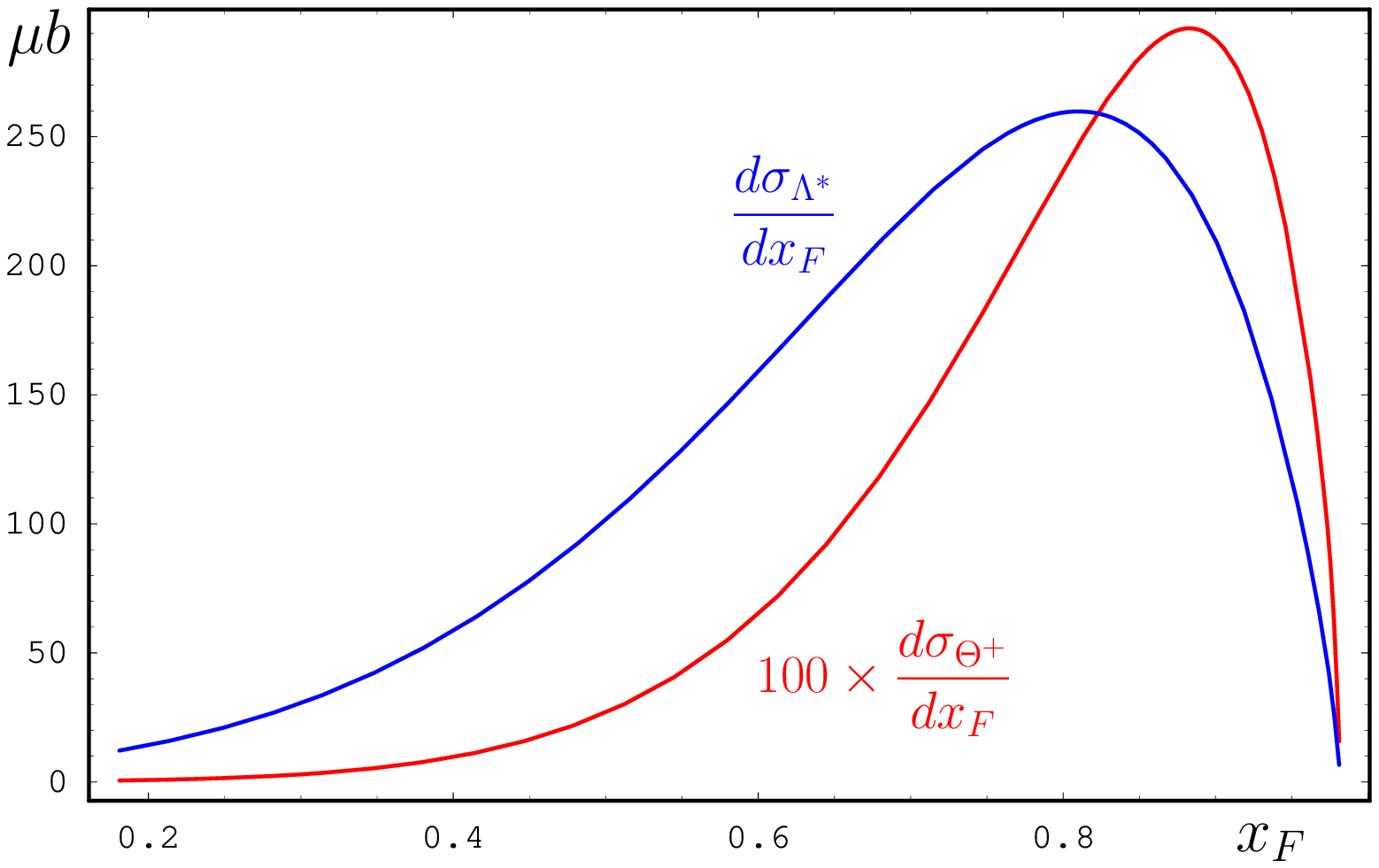}\hspace*{10mm}
\includegraphics[width=60mm,keepaspectratio=true]{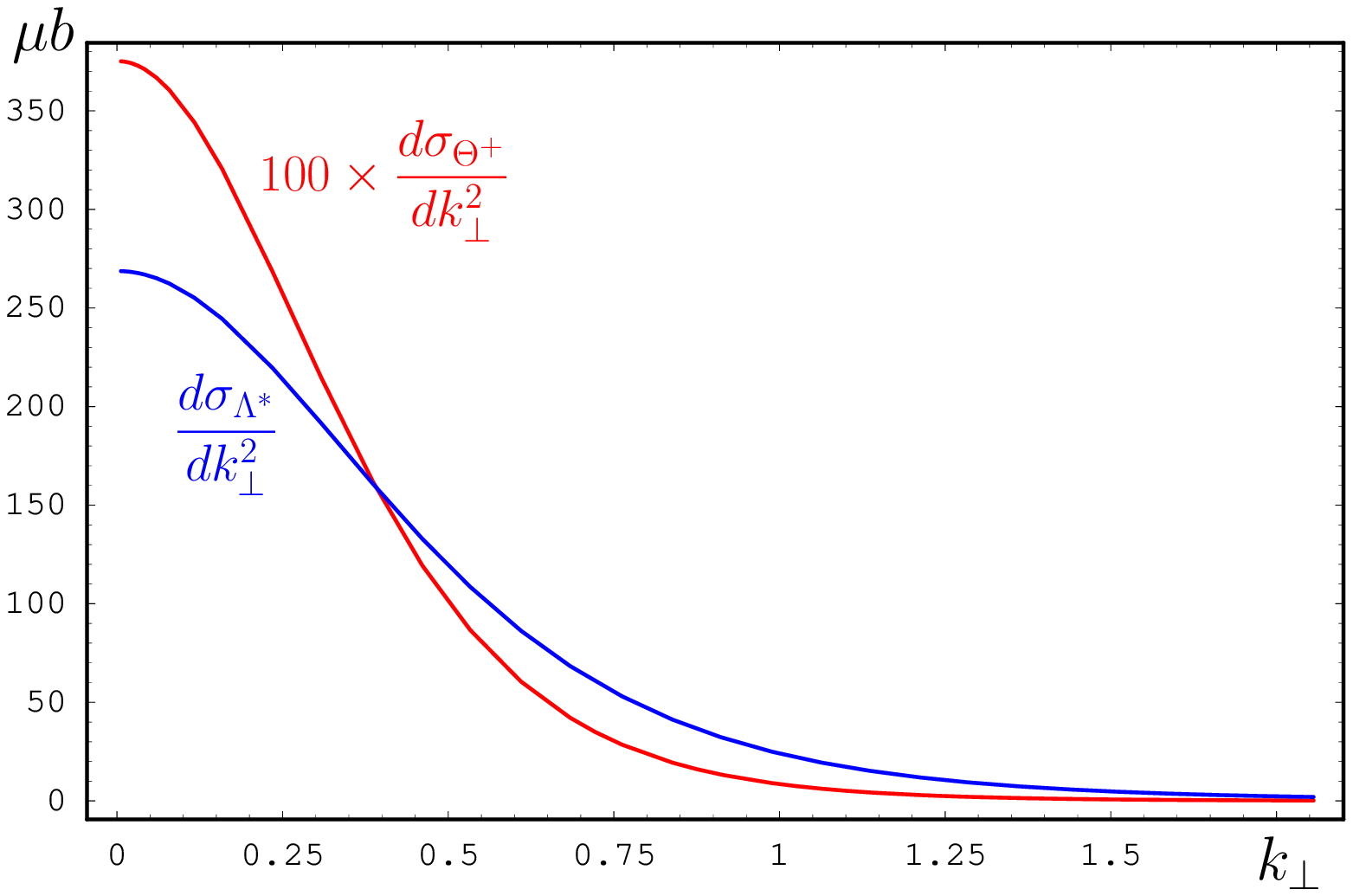}
\caption{$x_F$ (left) and $k_{\perp}$ (right) dependence of
inclusive cross sections} \label{dsigmadx}
\end{figure}
The inclusive cross sections can be obtained by integrating
(\ref{dsigma}) over $k_{\bot}^2$ and $x_F$ considering the total
cross sections $\sigma_{tot}(\bar K^0p)$ and $\sigma_{tot}(K^-p)$
as constants
%at the region $10~{\rm
%GeV}~\lesssim~\sqrt{s_1}~\lesssim~100~{\rm GeV}$. For estimations
%we take
($\sigma_{tot}(\bar K^0p)\sim\sigma_{tot}(K^-p)\sim 20~{\rm mb}$).
We take two representative examples for the form factor $F(t)$ in
(\ref{dsigma}):
%\begin{equation}
\begin{equation} \label{formfactors}F(t)=\frac{\Lambda^2-m_K^2}{\Lambda^2-t}~({\rm choice~A})~{\rm
and}~~F(t)=\frac{\Lambda^4}{\Lambda^4+(t-m_K^2)^2}~({\rm
choice~B})\end{equation}with~ $\Lambda=1~{\rm GeV}$.
%\end{equation}
Then we obtain for the
production cross sections
\begin{equation}
\sigma(pp\to \Theta^+X)= 0.8~(1.6)\times\frac{\Gamma_{\Theta
KN}}{1~{\rm MeV}}~\mu{\rm b},~~~~~\sigma(pp\to
\Lambda(1520)^+X)=106~(126)~\mu{\rm b},
\end{equation} where the first values refer to the form factor (A) in (\ref{formfactors}) and
the second ones to the form factor (B).
%\label{1}
%\end{equation}
The result for $\sigma(pp\to \Theta^+X)$ matches well that of Ref.
\cite{Vera} for the inclusive $pp\to\Theta^+X$ reaction at the
threshold and intermediate energies. One can expect that the $K^*$
exchange yields the cross sections of the similar order of
magnitude.

Our prediction for $\sigma(pp\to \Lambda(1520)^+X)$ agrees with
the preliminary result of the SVD-2 collaboration \cite{SVD}, but
$\sigma(pp\to \Theta^+X)$ is lower than the preliminary cross
section estimation (for $x_F>0$) of Ref. \cite{SVD}: $\sigma\cdot
{BR}(\Theta^+\to pK^0)\sim 6~\mu$b.

The $x_F$ distributions and transverse momentum distributions for
the $\Theta^+(1540)$ and $\Lambda(1520)$ are shown in Fig. 2. For
the average transfer momenta squared of $\Theta^+$ and
$\Lambda(1520)$ we get
\begin{equation} <k_{\bot}^2>~=~~0.25~{\rm GeV}^2~~~~{\rm
and}~~~~<k_{\bot}^2>~=~~0.25~{\rm GeV}^2,\end{equation}
respectively.

The ratio of $\Theta^+(1540)$ to $\Lambda(1520)$ production
cross-sections is $\sim 1\%$. Probably today our calculations can
be useful to explain why the $\Theta^+$ production is suppressed
in some high energy experiments.

\section*{Acknowledgements}

We are grateful to K.G.Boreskov, A.G.Dolgolenko, B.L.Ioffe and
A.B.Kaidalov for the discussions. This work was supported by RFBR
grants 03-02-17345, 04-02-17263, 05-02-17869, 05-02-27242 and by
the grant for leading scientific schools 1774.2003.2.


%\bibliography{sample}


\end{document}